\RequirePackage{fix-cm}
\documentclass[smallextended]{svjour3}       
\smartqed  
\usepackage{graphicx}
\begin{document}

\title{High resolution digitization system for the CROSS experiment
\thanks{The project CROSS is funded by the European Research Council (ERC) under the European Union Horizon 2020 program (H2020/2014-2020) with the ERC Advanced Grant no. 742345(ERC-2016-ADG).}
}


\author{P. Carniti$^{1,2,*}$		\and
        C. Gotti$^2$		\and
        G. Pessina$^2$		
}

\authorrunning{P. Carniti $\cdot$ C. Gotti $\cdot$ G. Pessina} 

\institute{$^1$ University of Milano Bicocca - Piazza della Scienza 3, 20126 Milan (Italy)\\
$^2$ INFN of Milano Bicocca - Piazza della Scienza 3, 20126 Milan (Italy)\\
$^*$ Corresponding author - \email{paolo.carniti@mib.infn.it}
}

\date{Received: date / Accepted: date}

\maketitle

\begin{abstract}
The signal digitization for CROSS, a bolometric experiment for the search of neutrinoless double beta decay at LSC – Canfranc Underground Laboratory, will be based on a custom solution comprised of an analog-to-digital board interfaced to an Altera Cyclone V FPGA module.
Each analog-to-digital board hosts 12 channels that allow data digitization up to 25~ksps per channel and an effective resolution of 21 bits at the typical sample rate required by the experiment (5~ksps).
The board also allows to digitally select the cut-off frequency of the anti-aliasing filter with 10 bit resolution from 24~Hz up to 2.5~kHz, as required by fast scintillating bolometers.
The FPGA is responsible for the synchronization of the analog-to-digital boards and for the data transfer to the storage, using UDP protocol on a standard Ethernet interface.
Each FPGA can manage the data coming from 8 boards (96 channels), allowing an excellent scalability.
In this contribution we will present a complete overview of the system, and a detailed characterization of the system performance.
\end{abstract}

\newpage

\section{Introduction}
\label{intro}
CROSS (Cryogenic Rare-event Observatory with Surface Sensitivity)~\cite{CROSS} is a bolometric experiment devoted to the search of neutrinoless double-$\beta$ decay that will be installed in Canfranc Underground Laboratory (LSC, Spain).

The main innovation of the CROSS experiment with respect to current generation of bolometric experiments, like CUORE~\cite{CUORE}, is the capability to discriminate bulk events from surface background events by using pulse-shape discrimination (PSD).
$\mathrm{0\nu2\beta}$ and $\mathrm{2\nu2\beta}$ candidate events, in fact, are expected to release their energy in the whole bulk of the crystal, while the main sources of background come from surface contamination of the crystal or from the surrounding environment.
The discrimination of surface events in CROSS is possible thanks to ultra-pure superconductive Aluminum thin foils, deposited on the surface of the crystals, which act as energy absorbers for $\alpha$ surface events, providing a different dynamic in the conversion and propagation of phonons, hence a different signal shape.
This technique has been already demonstrated both with Tellurium dioxide ($\mathrm{TeO_2}$)~\cite{AlCoating} and Lithium molybdate ($\mathrm{LiMoO_4}$)~\cite{CROSS} crystals.
The final choice of the $\mathrm{0\nu2\beta}$ candidate material is still under study.

For what concern the phonon sensors, two options are being considered: NTD Germanium thermistors (widely used in bolometric experiments) and NbSi thermistors (faster and more sensitive to athermal phonons).

The adoption of PSD techniques has important implications in the design of both the front-end and back-end electronics.
For what concern the DAQ system, there are some different requirements with respect to present $\mathrm{0\nu2\beta}$ bolometric experiments (CUORE~\cite{CUORE}, CUPID-0~\cite{CUPID0}, etc.):

\noindent\textit{Faster signals --} phonon signals in CROSS have rise times in the order of 1~ms (up to 500~Hz bandwidth), so the DAQ system should allow signal digitization at 5~ksps or faster.

\noindent\textit{Higher pile-up --} Lithium molybdate exhibits higher pile-up due to the higher $\mathrm{2\nu2\beta}$ background, and this also requires a fast sampling rate, in order to properly discriminate pile-up events.

\noindent\textit{Higher resolution --} detector noise will be reduced thanks to the adoption of a quieter cryostat setup, so $\sim$20 bit resolution will allow to exploit the large signal dynamic and the lower noise of both detector and front-end electronics.

\noindent\textit{Continuous acquisition --} this is required in order to apply more complex offline triggers and optimum filtering techniques, as successfully done in CUORE.

\noindent\textit{Spread in detector characteristics --} cut-off frequency should be finely tuned in order to adapt to the characteristics of each detector.

\section{DAQ board for CROSS experiment}
\label{sec:1}

\begin{figure}
	\centering
	\includegraphics[width=0.5\textwidth]{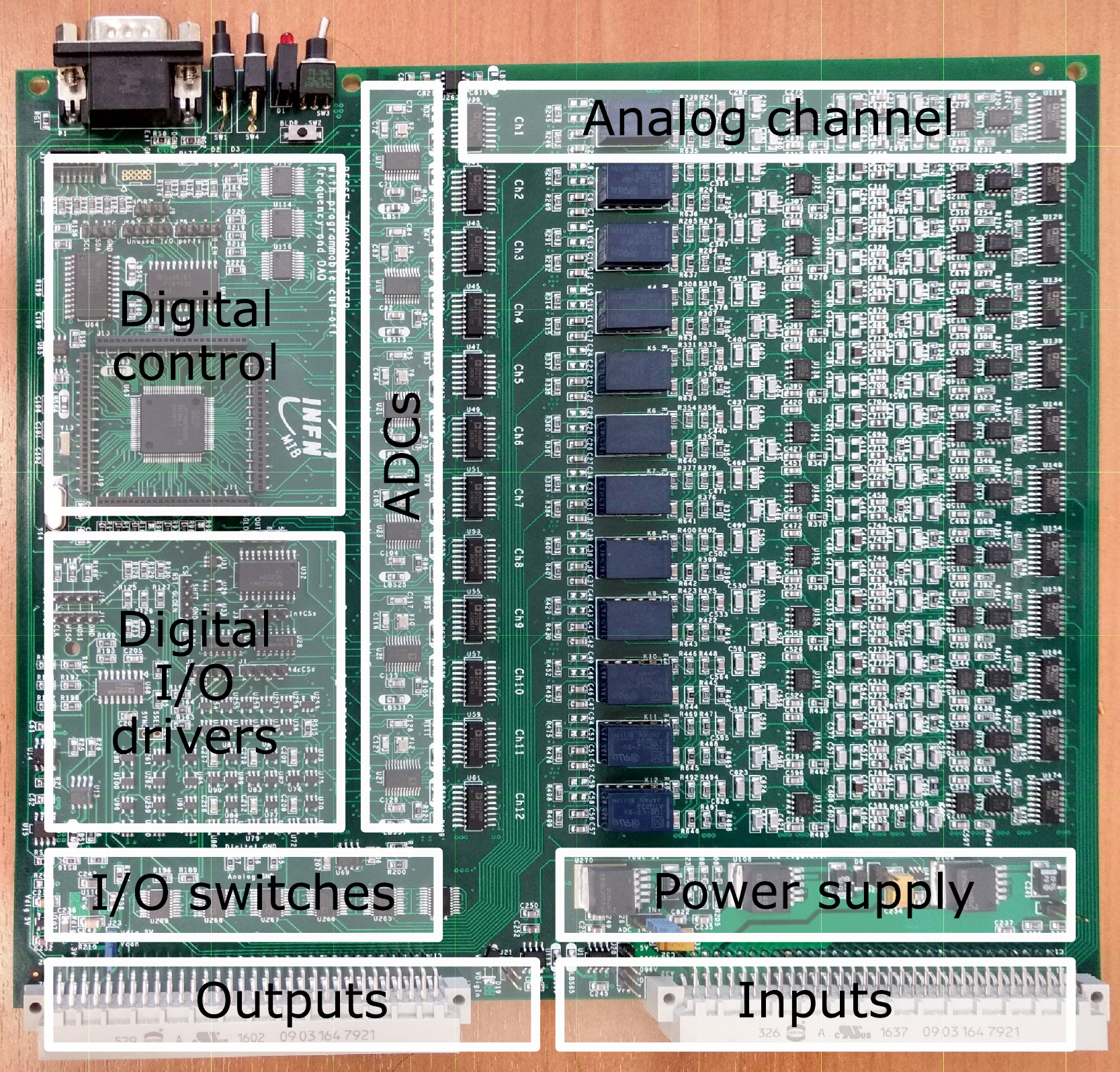}
	\caption{Photograph of the DAQ board for CROSS experiment with superimposed functional blocks.}
	\label{fig:scheda}       
\end{figure}

\begin{figure}
	\centering
	\includegraphics[width=0.75\textwidth]{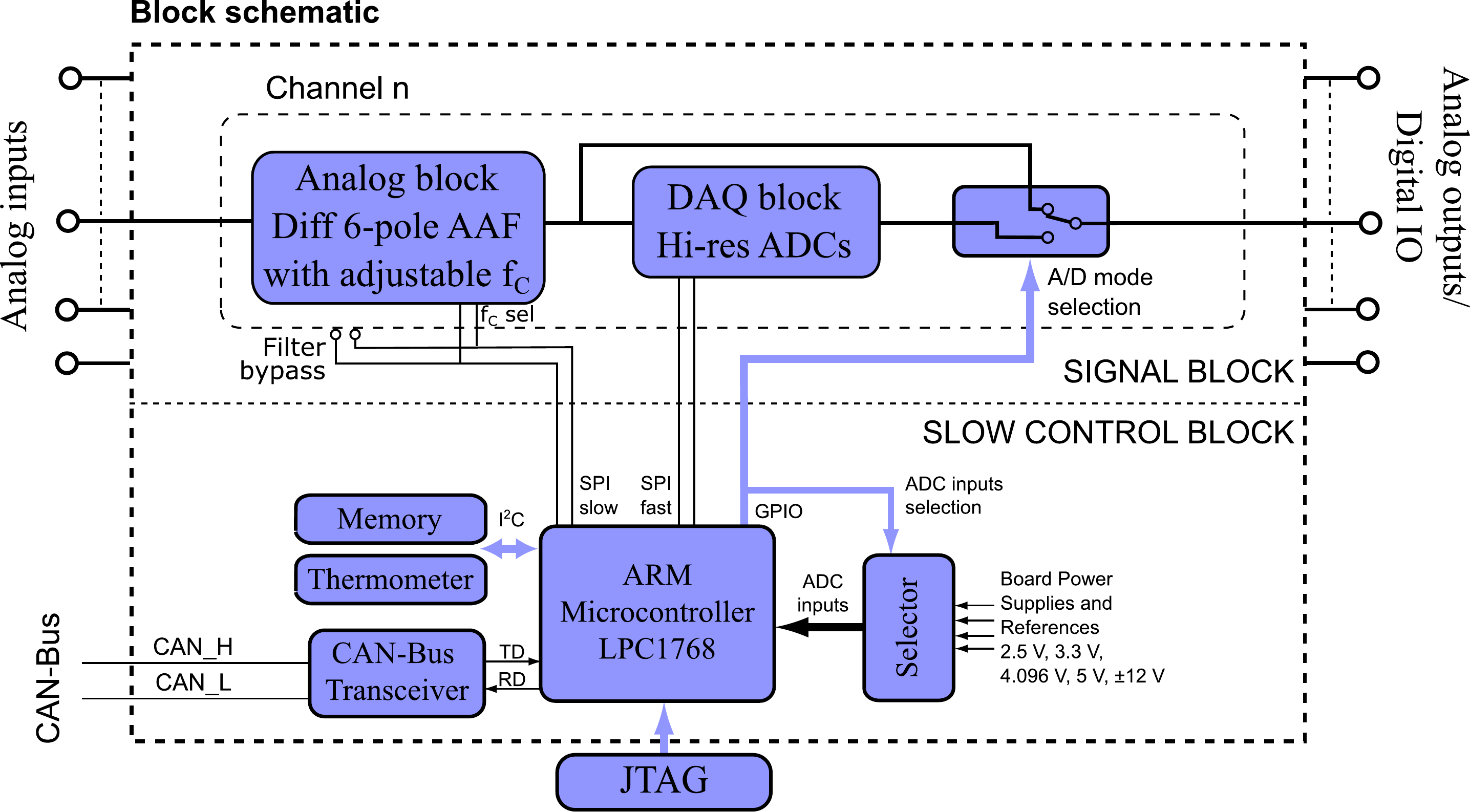}
	\caption{Block schematic of the DAQ board for CROSS.}
	\label{fig:schema}       
\end{figure}

The DAQ board that has been developed for the CROSS experiment is shown in Fig.~\ref{fig:scheda}, while Fig.~\ref{fig:schema} shows its block schematic.
The top part of the schematic shows the signal block for one channel out of the 12 available on the board.

\begin{figure}
	\centering
	\includegraphics[width=0.8\textwidth]{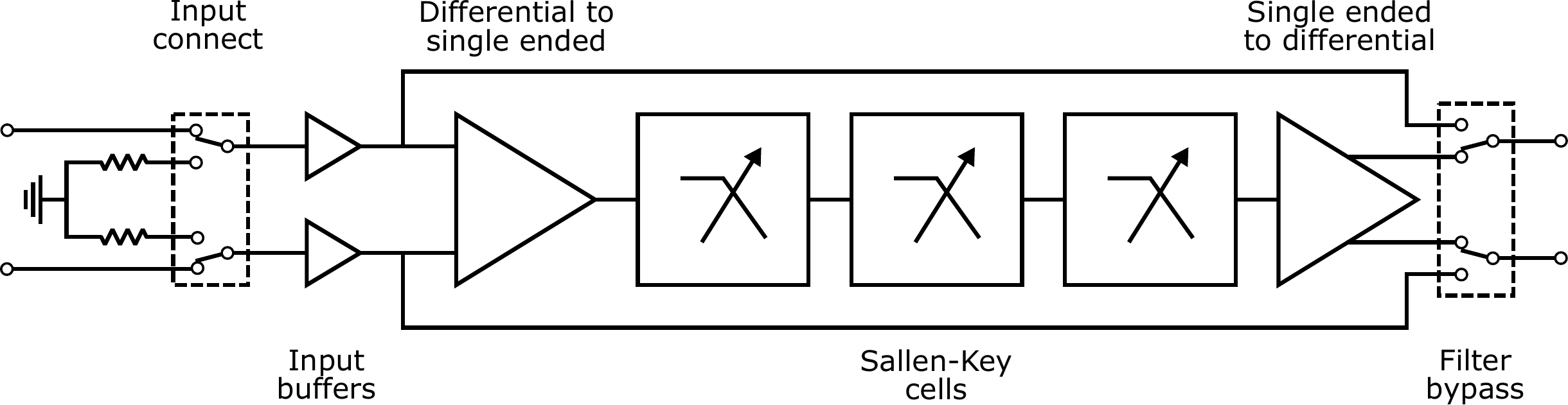}
	\caption{Analog filtering block schematic. Inputs are buffered and can be grounded. The low pass filter can be bypassed.}
	\label{fig:canale}       
\end{figure}

Each channel is equipped with a 6-pole Bessel-Thomson anti-aliasing filter with 10-bit digitally selectable cut-off frequency from 24~Hz up to 2.5~kHz using high precision digital trimmers (1\% absolute accuracy, $\mathrm{35\ ppm/^{\circ}C}$ drift). The trimmers have a resistance of $\mathrm{100\ k\Omega}$, which allows to use high quality C0G capacitors of few tens of nF. 
Fig..~\ref{fig:canale} shows in more detail the schematic of the analog block.
Inputs are buffered and can be grounded to allow offset self-calibration.
The anti-aliasing filter can be bypassed.

Each pair of analog channels is equipped with dual-channel 24-bit $\mathrm{\Delta\Sigma}$ ADCs which are able to digitize the signals up to 25~ksps/ch in 12-channel mode or 250~ksps/ch in 6-channel mode.

The board has output switches that allow it to operate in fully analog mode, with filtered analog signals provided to the external connector, or in fully digital mode, with digital lines to the ADCs provided to the external connector.
In the CROSS experiment, this latter configuration will be used.
The bottom part of the block schematic (Fig.~\ref{fig:schema}) shows the slow control circuitry, responsible of the setting of all the onboard peripherals, as well as monitoring.
The onboard slow control circuit is connected via CAN bus to the experiment slow control server.

\begin{figure}
	\centering
	\includegraphics[width=0.55\textwidth]{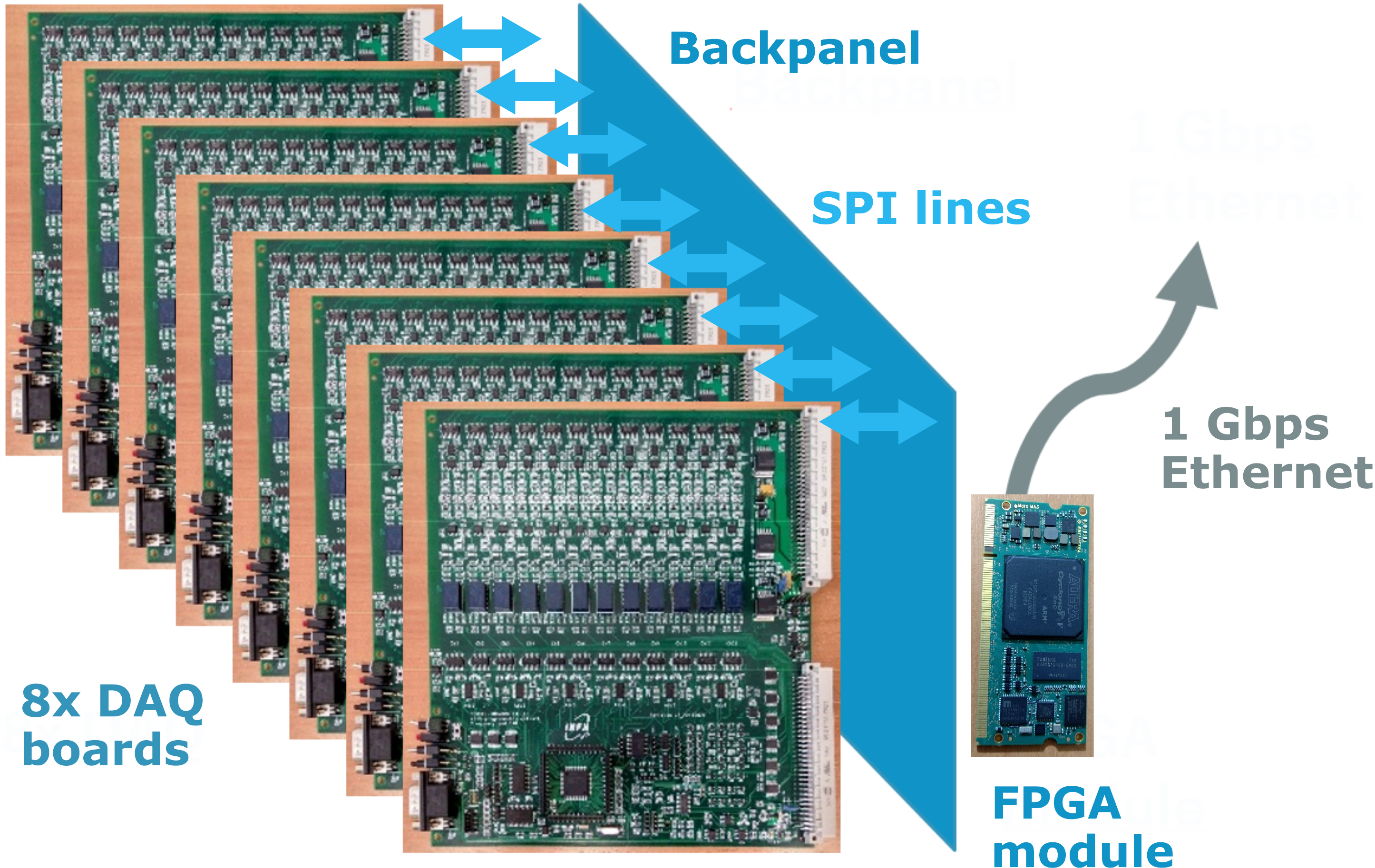}
	\caption{Back-end block schematic. The FPGA is installed on the backplane and collects data from 8 boards through high speed (20~MHz) SPI lines.}
	\label{fig:backend}       
\end{figure}

When operating in digital mode, as in CROSS, the digitized signals from the on-board ADCs are extracted from the ADCs by a commercial FPGA module (Enclustra Mars MA3) installed on the backpanel that collects the data from 8 boards (96 channels).
The data interface to the ADCs uses 6 high speed (20~MHz) SPI lines, while slow control of the board is mediated by an on-board ARM Cortex-M3 microcontroller interfaced to the FPGA with CAN bus.
This data is then transferred to the storage system using an inexpensive 1~Gbps Ethernet interface (optically decoupled).
The chosen data transfer protocol is UDP (hardware-synthesized in the FPGA), with a maximum data rate of 768~Mbps (6 channels per board at 250~ksps and 64 bit data length).
An illustration of the back-end communication is shown in Fig~\ref{fig:backend}.

\section{Board performance}
\label{sec:2}

The first samples of the board have been fully characterized in order to evaluate their performance.
A summary of the board specifications and test results are gathered in Table~\ref{tab:spec}.

\begin{table}
\centering
\caption{Summary of specifications and performance of the system.}
\label{tab:spec}       
\includegraphics[width=0.6\textwidth]{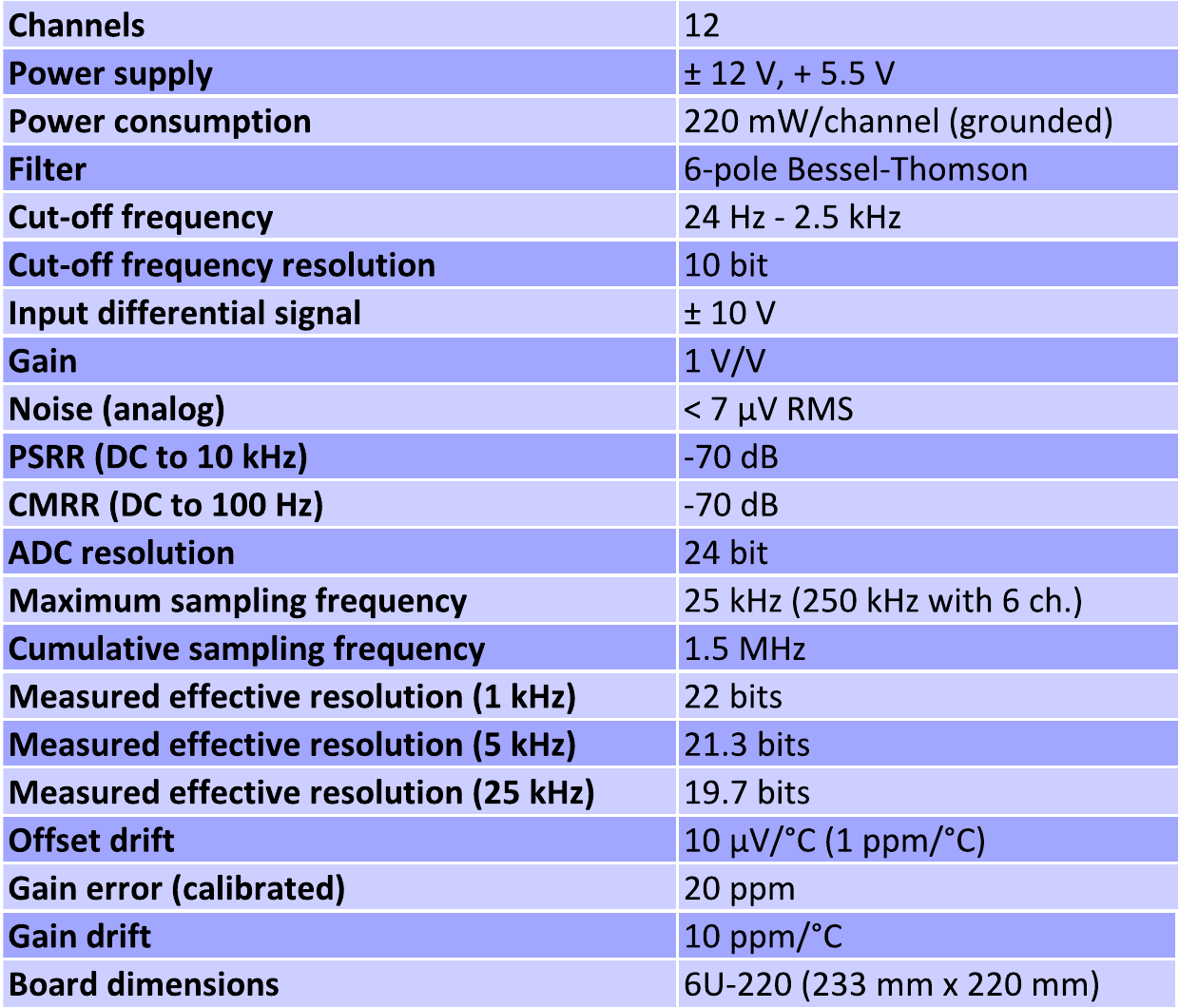}
\end{table}

\begin{figure}
	\centering
	\includegraphics[width=0.45\textwidth]{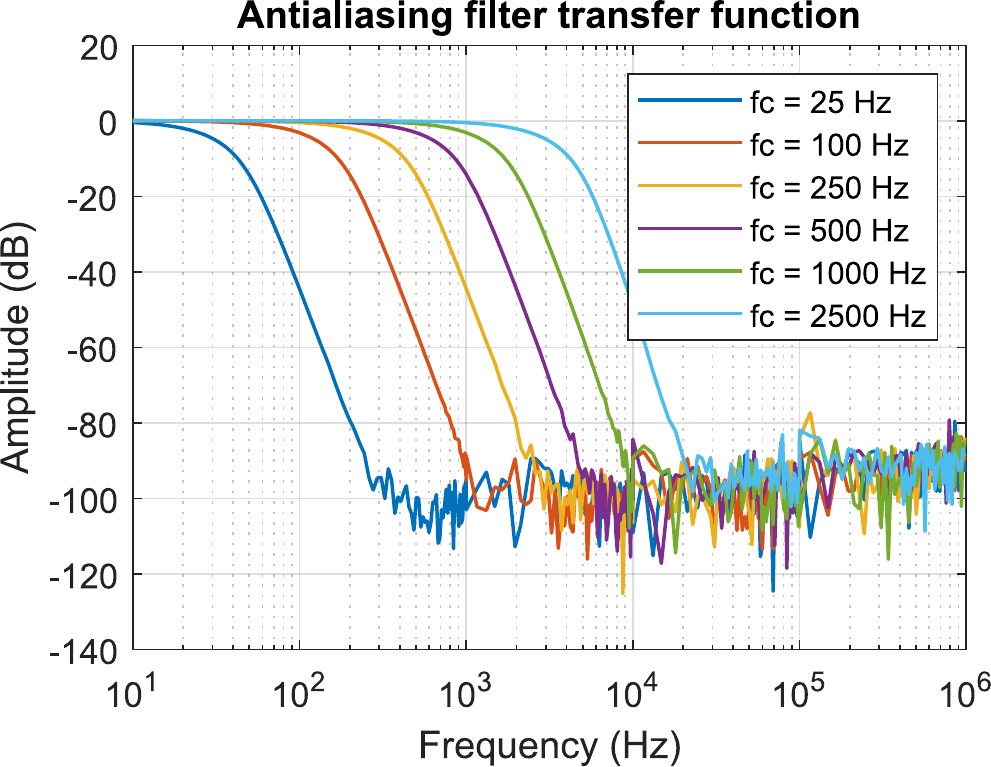}
	\caption{Analog filter transfer function at different cut-off frequency settings.}
	\label{fig:filter}       
\end{figure}

The plot in Fig.~\ref{fig:filter} shows the adjustment of the cut-off frequency using the digital trimmer.
Six settings between 25~Hz and 2.5~kHz are selected and the transfer function is measured using an Agilent 4395A spectrum analyzer.

\begin{figure}
	\centering
	\includegraphics[width=0.5\textwidth]{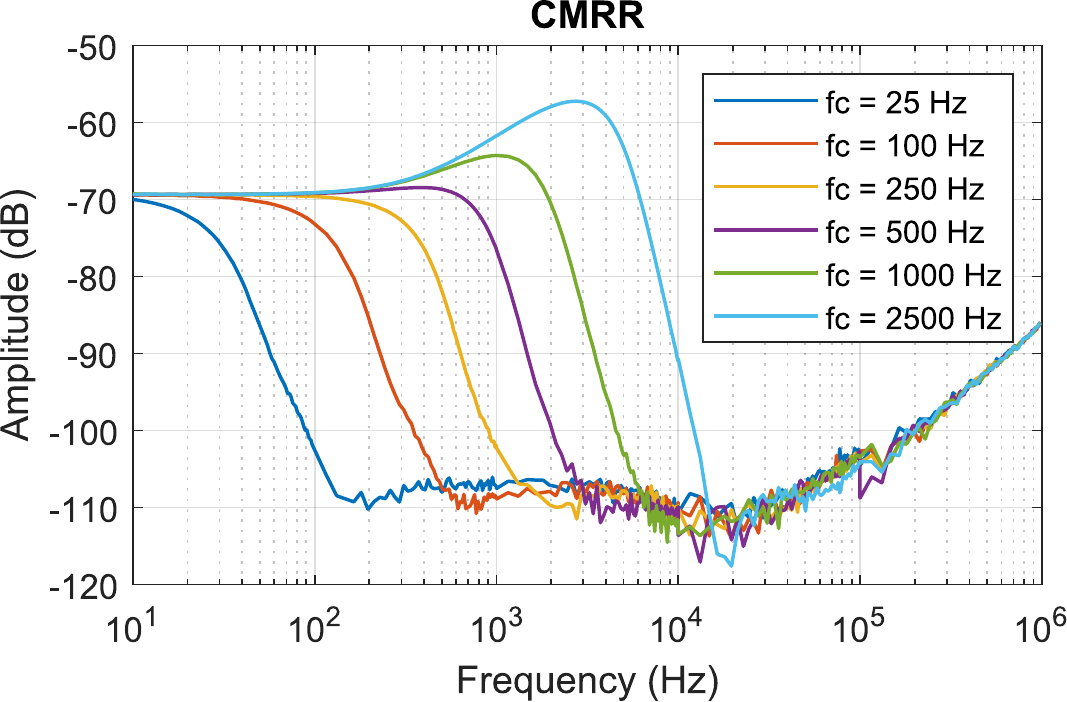}
	\caption{Common mode rejection ratio (CMRR) at different cut-off frequency settings.}
	\label{fig:cmrr}       
\end{figure}

The common mode rejection ratio (CMRR) is plotted in Fig.~\ref{fig:cmrr}.
In DC, the CMRR is about $\mathrm{-70\ dB}$, mainly dominated by the mismatch in the gain of the input buffers (i.e. resistor mismatches). This value is largely sufficient for this application where a front-end stage is present.

\begin{figure}
	\centering
	\includegraphics[width=0.5\textwidth]{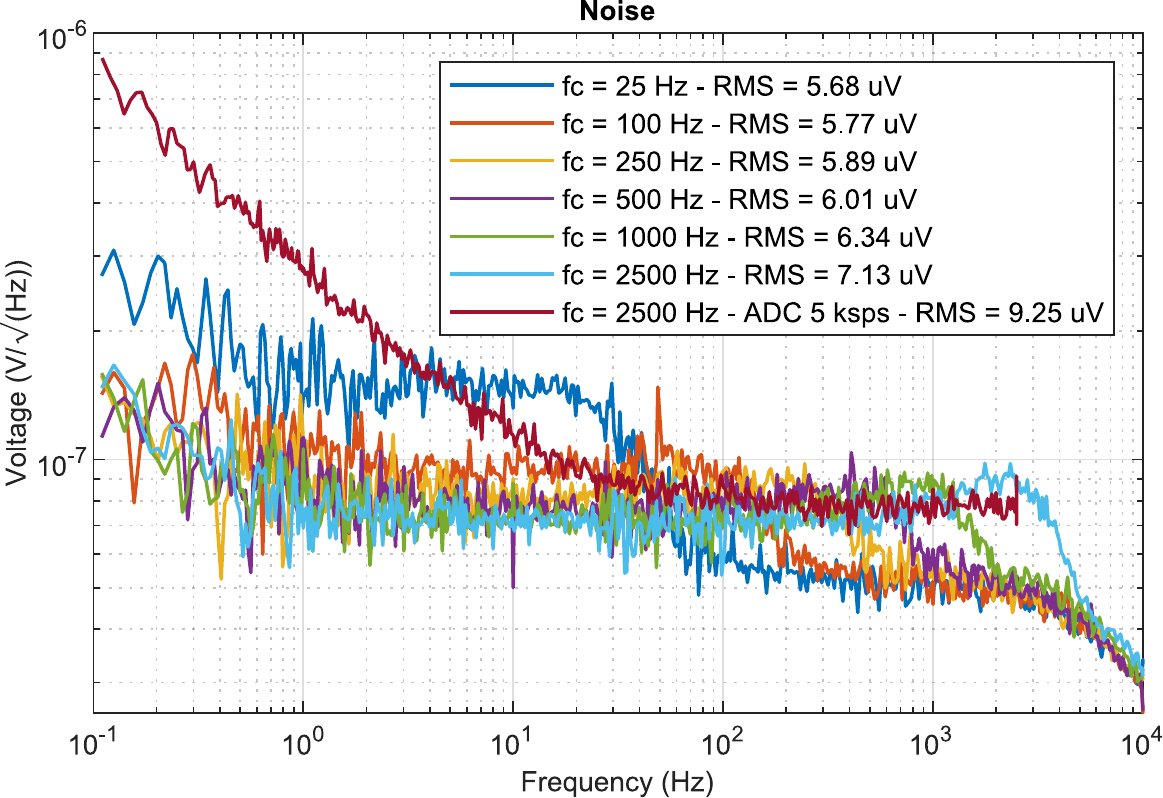}
	\caption{Spectra of the analog noise at different cut-off frequency settings. As comparison, the noise spectrum of the digitized data at 5 ksps is also shown.}
	\label{fig:noise}       
\end{figure}

Fig.~\ref{fig:noise} shows the noise of the analog filter at the same cut-off frequency settings.
At low filter bandwidth ($<100$~Hz), noise is dominated by the series noise of the trimmers ($\mathrm{100\ k\Omega}$ at 24~Hz), while at higher filter bandwidth, series noise of operational amplifiers becomes predominant.
The contribution of both parallel and 1/f noise of the opamps was made negligible with proper selection of the opamps themselves.
The total analog RMS noise is indicated in the legend of Fig.~\ref{fig:noise} and ranges from 5.7~$\mathrm{\mu}$V at 25~Hz up to 7.1~$\mathrm{\mu}$V at 2.5~kHz (0.1~Hz -- 100~kHz bandwidth).

When the signals are read-out using the internal $\mathrm{\Delta\Sigma}$ ADCs, the effective resolution is dominated by the noise of the ADC and its buffer at sampling frequencies above 5~ksps.
The effective resolution is 22.0 bits at 1~ksps, 21.1 bits at 5~ksps and 19.7 bits at 25~ksps, which correspond to 4.7~$\mathrm{\mu}$V, 9.2~$\mathrm{\mu}$V and 24~$\mathrm{\mu}$V RMS respectively.
A noise spectrum of digitized data at 5 ksps is shown in Fig.~\ref{fig:noise}.

The board has also been qualified in a climatic chamber in order to measure its stability against temperature variations, which is very important in this kind of experiments that are expected to run for long times while monitoring detector baseline stability. Offset drift was measured to be $\mathrm{10\ \mu V/^{\circ}C}$ ($\mathrm{1\ ppm/^{\circ}C}$), while gain sensitivity was measured to be $\mathrm{10\ ppm/^{\circ}C}$.

\section{Conclusions and future developments}
\label{sec:3}

The board is already fully operational and characterized.
It offers many improvements over widely used commercial solutions, mainly for what concern noise, stability and power consumption.
Data read-out from the FPGA module has been successfully demonstrated using a provisional backpanel.
Using this setup, a test with prototype CROSS detectors is foreseen in the near future, in order to qualify the system with real data from the bolometers and to compare its performance with standard commercial solutions.
Following this test, the full backpanel will be designed, which will allow to test the system at full output data rate with multiple DAQ boards and also implement the required synchronization between different FPGA modules.


\begin{thebibliography}{}
%
%

\bibitem{CROSS}
I. C. Bandac et al., \textit{The $0\nu2\beta$-decay CROSS experiment: preliminary results and prospects}, arXiv:1906.10233 (2019)
\bibitem{CUORE}
C. Alduino, \textit{First Results from CUORE: A Search for Lepton Number Violation via ${0\nu\beta\beta}$ Decay of ${^{130}Te}$}, Phys. Rev. Lett. 120 (2018) 132501, DOI: 10.1103/PhysRevLett.120.132501
\bibitem{AlCoating}
C. Nones et al., \textit{Superconducting Aluminum Layers as Pulse Shape Modifiers: An Innovative Solution to Fight Against Surface Background in Neutrinoless Double Beta Decay Experiments}, J. Low Temp. Phys. 167 (2012) 1029–1034, DOI: 10.1007/s10909-012-0558-y.
\bibitem{CUPID0}
O. Azzolini, \textit{First Result on the Neutrinoless Double-${\beta}$ Decay of ${^{82}Se}$ with CUPID-0}, Phys. Rev. Lett. 120 (2018) 232502, DOI: 10.1103/PhysRevLett.120.232502

\end{thebibliography}
\end{document}